\documentclass[aps,prx,twocolumn,
showpacs,showkeys,
nobalancelastpage,
nofootinbib,
longbibliography
]{revtex4-1}
\usepackage{graphicx}
\usepackage{indentfirst}
\usepackage{physics}
\usepackage{braket}
\usepackage{float}
\usepackage{amsmath}
\usepackage{amssymb}
\usepackage{verbatim}
\usepackage{epstopdf}
\usepackage{CJK}
\usepackage{esint}
\usepackage{color}
\usepackage{epsfig}
\usepackage{subfigure}
\usepackage{amsfonts}
\usepackage{footmisc}
\usepackage{scrextend}
\usepackage{multirow}
\usepackage{mathtools}
\usepackage{xr-hyper}
\usepackage[hyperfootnotes=false]{hyperref}

\usepackage{tikz,pgfplots}
\usepackage{mathdots}
\usepackage{yhmath}
\usepackage{cancel}
\usepackage{scalerel}
\usepackage{color}
\usepackage{siunitx}
\usepackage{array}
\usepackage{multirow}
\usepackage{amssymb}
\usepackage{gensymb}
\usepackage{tabularx}
\usepackage{booktabs}
\usetikzlibrary{fadings}
\usetikzlibrary{patterns}
\usetikzlibrary{shadows.blur}

\usepackage[english]{babel}
\usepackage{url}
\usepackage{bm}
\definecolor{darkblue}{rgb}{0,0,0.5}
\hypersetup{
colorlinks=true,
linkcolor=black,
filecolor=black,
citecolor=darkblue,
urlcolor=darkblue,
}

\newcommand{\defeq}{\vcentcolon=}

\newcommand\mc[1]{\mathcal{#1}}
\newcommand\bs[1]{\boldsymbol{#1}}

\begin{document}

\title{Idler-Free Multi-Channel Discrimination via Multipartite Probe States}
\author{Cillian Harney}
\email{cth528@york.ac.uk}
\author{Stefano Pirandola} 
\email{stefano.pirandola@york.ac.uk}
\affiliation{Department of Computer Science, University of York, York YO10 5GH, United Kingdom}
\begin{abstract}
The characterisation of Quantum Channel Discrimination (QCD) offers critical insight for future quantum technologies in quantum metrology, sensing and communications. The task of multi-channel discrimination creates a scenario in which the discrimination of multiple quantum channels can be equated to the idea of pattern recognition, highly relevant to the tasks of quantum reading, illumination and more. 
Whilst the optimal quantum strategy for many scenarios is an entangled idler-assisted protocol, the extension to a multi-hypothesis setting invites the exploration of discrimination strategies based on unassisted, multipartite probe states. In this work, we expand the space of possible quantum enhanced protocols by formulating general classes of unassisted multi-channel discrimination protocols which are not assisted by idler modes. 
Developing a general framework for idler-free protocols, we perform an explicit investigation in the bosonic setting, studying prominent Gaussian channel discrimination problems for real world applications. Our findings uncover the existence of strongly quantum advantageous, idler-free protocols for the discrimination of bosonic loss and environmental noise. This circumvents the necessity for idler assistance to achieve quantum advantage in some of the most relevant discrimination settings, significantly loosening practical requirements for prominent quantum sensing applications.
\end{abstract}
\maketitle

\section*{Introduction}

As the development of practical quantum technologies accelerates \cite{Mike_Ike, preskillnisq, GoogleSyc}, the field of quantum sensing is already the most mature, and already obtaining quantum advantage in a variety of applications \cite{QSensing}. 
Critical theoretical underpinnings in quantum metrology and hypothesis testing \cite{HelstromQHT, BaeQSD, QSDBergou, QSDChefles, DiscrimQOps,QMetr} have led to quantum enhanced protocols with fundamental applications in quantum illumination \cite{Lloyd2008, Tan2008, Shapiro2009, Zhang2013, Zhang2015, Barzanjeh2015, Dallarno2012, Zhuang2017_1, Zhuang2017_2, LasHeras2017, DePalma2018, NairIllum,Lopaeva2013, Barzanjeh2020} and quantum reading \cite{PirBD, NairNDS,Tej2013, Spedalieri2012, GSCryptoQR, Zhuang2017_3,Hirota2017, expqread,QEnhDataCl}, with particular interest in the Continuous Variable (CV) domain \cite{GaussRev, SerafiniCV, AdessoGInfo}. \par

The fundamental task of Quantum Channel Discrimination (QCD) models many of these applications. In QCD, a user is tasked with classifying an ensemble of quantum channels through the use of an input quantum state (probe state) and a discriminatory measurement. Locating an optimal discrimination protocol is very difficult, as it embodies a double optimisation problem of both the probe state and the output measurement. Nonetheless, significant progress has been made in recent years in a variety of contexts \cite{FLQCD, AmpDampB}.\par

Until recently, QCD has been mostly limited to the problem of binary classification. However, advances in multi-channel discrimination and the formulation of Channel Position Finding (CPF) \cite{EntEnhanced, ULMCD} have brought with them new insight and opportunities for more complex multi-hypothesis classification problems. These multi-channel discrimination problems are highly relevant in a number of fascinating settings, such as data-readout from optical memories, quantum enhanced optical/thermal pattern recognition \cite{PatternRecog, ThermalPatt}, and target detection \cite{OptEnvLoc}.

Within these applications (and many more in quantum sensing) the assistance of idler modes has been a crucial feature in order to attain quantum enhanced performance \cite{AdvPhS, EntEnhanced}. Idler-modes refer to perfectly preserved, ancillary quantum systems which share entanglement with input probe states throughout a sensing protocol. In the bosonic setting, these protocols consist of using one mode of a Two Mode Squeezed Vacuum (TMSV) state to probe a target, whilst the remaining mode (the idler) is kept by the user. Idler-assisted protocols have been shown to be optimal for a number of important discrimination tasks, and offer significant advantage for many more. \par

Yet, the necessity for idler-modes is problematic, due to the requirement that they need to be perfectly protected. In practice this is not possible, as some decoherence will always be imparted on the idler while the probe mode is interacting with a target. To combat this, idlers are either contained in delay lines (e.g.~very low-loss fibre optics) or stored in quantum memories until required for measurement. This preservation requirement causes serious practical difficulties due to the challenging nature of creating stable quantum memories with adequate storage time \cite{Lvovsky_QMem_2009, Jensen_QMem_2010,Cho_QMem_2016,Wang_QMem_2019}. In some settings, it may be much more practical to use unassisted protocols limited to signal-only probe modes, especially for near term quantum technologies. \par

Research on unassisted protocols has been primarily limited to single channel sensing problems, motivating the use of coherent states to formulate classical benchmarks, and even to search for quantum enhancements beyond entanglement \cite{BarelySep, QSBE}. However the multi-channel discrimination picture invites us to explore different unassisted protocols. 
In particular, it is now possible to construct protocols that distribute entanglement over multiple quantum channels using multipartite entangled states. Without additional idler modes to defend entanglement, input states must be cleverly designed to preserve quantum correlations in the face of increased decoherence.
Recently, Pereira et al.~\cite{Jason_IdlerFree} have explored the use of a block protocol with entangled bosonic states for discriminating small collections of Gaussian quantum channels, proving that there do exist idler-free protocols capable of exceeding the best known classical strategies. 

Motivated by this, we arrive at our key research objectives: To generalise the theory of unassisted protocols for multi-channel discrimination, and to ask: Can we design unassisted multi-channel discrimination protocols that achieve significant quantum advantage?

Hence, in this work, we construct general classes of unassisted protocols for multi-channel discrimination. These are block protocols which utilise (generally entangled) multipartite quantum states as probe states. Multipartite states (and thus entanglement) can now be distributed across multiple quantum channels in many inequivalent ways, leading to two distinct, broad classes of discrimination protocols. Via multi-mode entanglement and carefully designed probe distributions, we present unassisted protocols that are able to attain performances on par with that of idler-assistance. This circumvents the necessity for idler-assistance in some of the most relevant discrimination settings, loosening practical requirements for quantum-enhanced pattern recognition.

This paper is structured as follows: In Results, after first reviewing the model of quantum pattern recognition, we present our main findings. We introduce the general framework of block protocols using unassisted multipartite quantum probe states. We then identify two distinct classes of unassisted protocol which emerge from this framework, discuss their operational interpretations, and devise a diagrammatic language for describing such protocols. We corroborate these general findings by demonstrating the efficacy of idler-free protocols for the discrimination of multiple bosonic Gaussian quantum channels. In Discussion, our results are summarised and we identify future investigative paths. Finally, the Methods section contains a number of useful theoretical tools and insights used within this research.

\section*{Results\label{sec:PSet}}

\subsection*{Quantum Pattern Recognition}
In this work, we study the discrimination of quantum multi-channels which we call \textit{quantum channel patterns}. 
A binary channel pattern is defined as an $m$-length sequence of quantum channels, such that each channel in the sequence admits the properties of a target channel $\mc{E}_T$ or background channel $\mc{E}_B$ (identified by the labels $T, B$ respectively). It is useful to convert this sequence into a multi-set of binary variables which represents the channel pattern $\bs{i} = \{i_1, i_2, \ldots, i_m\}$, where $i_j \in \{B,T\}$ for all $j \in \{1,\ldots,m\}$. We can then more precisely denote an $m$-length channel pattern as tensor product
\begin{align}
\mc{E}_{\bs{i}} &\defeq \mc{E}_{i_1} \otimes \mc{E}_{i_2} \otimes \ldots \otimes \mc{E}_{i_m} = \bigotimes_{j=1}^m \mc{E}_{i_j}.
\end{align}
Throughout this work, we refer to a channel pattern simply by its binary string $\bs{i}$, unless $\mc{E}_{\bs{i}}$ is formally required. Background and target channels can be used to encode physical properties of a multipartite system. For instance, one can associate each pixel of an $m$-pixel binary thermal image with a cold (background) or hot (target) temperature. Quantum mechanically, one may attribute each pixel to a quantum channel that describes how a quantum probe may interact with either pixel.\par 

A channel pattern $\bs{i}$ represents only a single instance of a possible binary arrangement. More generally, these instances belong to a larger space of multi-channels we may call an \textit{image space}. We label an arbitrary $N$-element image space as the set $\mc{U} = \{\bs{i}_1, \bs{i}_2, \ldots, \bs{i}_N\}$ containing $N$ unique channel patterns. Since we are considering binary patterns, the most general image space we can consider is the set of all $m$-length binary strings ${\mc{U}_{m} = \{\bs{i}_1, \bs{i}_2, \ldots, \bs{i}_{2^m}\}}$, of which all other binary image spaces are a subset. Image spaces can be used to specify important, physical problem settings such as those defined by the task of Channel Position Finding (CPF), which is concerned with locating target channels hidden amongst collections of background channels (see Methods for more details). \par

The challenge of multi-channel discrimination may now be presented: Consider an $m$-length pattern of unidentified quantum channels. Suppose that the sequence of channels belongs to a pattern from a known image space $\mc{U}$. Each pattern in the image space possesses a unique probability of existing, $\pi_{\bs{i}}$. The task of discrimination then consists of distinguishing between all the multi-channels in the statistical ensemble $\{ \pi_{\bs{i}} ; \mc{E}_{\bs{i}} \}_{\bs{i}\in\mc{U}}$, which describes an ensemble of multi-channels $\{\mc{E}_{\bs{i}} \}_{\bs{i}\in \mc{U}}$ distributed according to the classical probability distribution $\{\pi_{\bs{i}} \}_{\bs{i}\in \mc{U}}$.\par

The most general multi-channel discrimination protocol is a general adaptive protocol, $\mathcal{P}$. This is best described by a quantum comb \cite{GutoskiComb, ChiribellaComb,Laurenza2018}; a quantum circuit board with an arbitrary number of registers, with $M$ slots in which channel patterns $\mathcal{E}_{\bs{i}}$ are placed. There is no limit to the amount of entanglement that can used to construct a quantum comb, and a general adaptive protocol can make use of adaptive operations and feedback based state preparation. Due to their generality, these protocols are very difficult to characterise and optimise. Therefore it is often much more beneficial to consider simpler protocols. \par
Of such, block protocols $\mathcal{B}$ represent a very important class of non-adaptive discrimination strategy. Channel patterns are probed using $M$ identical and independent copies of some input probe state, ${\rho^{\otimes M} \rightarrow \rho_{\bs{i}}^{\otimes M} \defeq  \mathcal{E}_{\bs{i}}(\rho)^{\otimes M}}$.
After $M$ pattern interactions, an optimised POVM $\{\Pi_{\bs{i}}\}_{\bs{i} \in \mathcal{U}}$ is used to perform the classification. Given an image space $\mc{U}$ with the pattern probability distribution $\{\pi_{\bs{i}}\}_{{\bs{i} \in \mathcal{U}}}$, the average error probability of misclassification is given by
\begin{equation}
p_{\text{err}} (\mc{B}) \defeq  \sum_{\bs{i} \neq \bs{i}^{\prime} \in \mc{U}} \pi_{\bs{i}} \text{Tr}\left[ {\Pi}_{\bs{i}^{\prime}} \rho_{\bs{i}}^{\otimes M}\right] , \label{eq:p_err_block}
\end{equation}
where this sum runs over all pairs of unequal channel patterns throughout the image space. In order to benchmark this discrimination performance without specifying precise measurements, the following fidelity based bounds from can be used \cite{LBMont, UBPGM},
\begin{align}
p_{\text{err}} &\geq \frac{1}{2} \sum_{\bs{i}\neq\bs{i}^{\prime} \in \mc{U}} \pi_{\bs{i}} \pi_{\bs{i}^{\prime}} F^{2M}(\rho_{\bs{i}},\rho_{\bs{i}^{\prime}}), \label{eq:LB}\\
p_{\text{err}} &\leq \sum_{\bs{i}\neq\bs{i}^{\prime} \in \mc{U}} \sqrt{\pi_{\bs{i}} \pi_{\bs{i}^{\prime}}} F^{M}(\rho_{\bs{i}},\rho_{\bs{i}^{\prime}}), \label{eq:UB}
\end{align}
where $F(\rho,\sigma) = \text{Tr}\left[ \sqrt{\sqrt{\rho} \>\sigma \sqrt{\rho}}\right]$ denotes the Bures fidelity. These bounds are completely general, and do not depend on the channel dimension. Hence, they may be utilised for both finite and infinite dimensional input states (provided that we use energy-constrained quantum states).\par

These non-adaptive block protocols have been shown to offer high performance in a number of discrimination settings, and in some cases are optimal \cite{FLQCD}. If a block protocol makes use of entangled, ancillary quantum systems (idlers) then it is known as a block-assisted protocol $\mc{B}^{\text{a}}$. Idler-based entanglement can induce quantum-enhancements in many different discrimination settings \cite{PatternRecog,ThermalPatt}. Without additional idler modes, we are left with an unassisted block protocol, $\mc{B}^{\text{u}}$. Much less is known about unassisted protocols in a multi-channel setting, which we rectify in the following sections.

\subsection*{Fixed Unassisted Block Protocols}

Consider an image space $\bs{i} \in \mc{U}$ of $m$-length multi-channels each of which occur with probability $\pi_{\bs{i}}$, generating the channel pattern ensemble $\{\pi_{\bs{i}} ; \mc{E}_{\bs{i}} \}_{\bs{i}\in\mc{U}}$. Unassisted discrimination involves developing a strategy for accurately distinguishing patterns from the image space without utilising entangled idler-modes or ancillary quantum systems. Unlike in an assisted protocol, entanglement is now only permitted between probe modes. We proceed in this practical direction by investigating how inter-probe entanglement can play a role in constructing quantum enhanced, unassisted block protocols.\par

Consider an $m$-length channel pattern. An unassisted block protocol $\mc{B}^{\text{u}}$ using multipartite states will assign an $M$-copy, $n \leq m$ multi-mode state to interact with some region of the channel pattern, defined by a set of channel labels $\bs{s} = \{s_1, \ldots, s_n\}$ for $s_i \in \{1,\ldots,m\}$. This channel region $\bs{s}$, which we aptly call a \textit{probe-domain}, defines a sub-pattern of the total channel pattern over which a multipartite state $\sigma_{\bs{s}}^{\otimes M}$ can be irradiated. Hence, a probe-domain $\bs{s}$ is a sub-set of channel labels $\bs{s} \subseteq \{1,\ldots,m\}$ which designates a region of the channel pattern over which probe modes are permitted to be entangled. Input modes which are incident in the domain $\bs{s}$ can be entangled but are fully separable with respect to any modes outside of this region. Furthermore, each $M$-copy probe state $\sigma_{\bs{s}}^{\otimes M}$ can be different for its respective probe-domain.\par

In order to completely interact with all $m$-channels in the pattern it is necessary to define a discrete \textit{probe-domain distribution}. This is a collection of distinct channel pattern sub-regions $\{ \bs{s}_1, \bs{s}_2, \ldots, \bs{s}_N\}$ over which an associated $N$ length collection $M$-copy multipartite states $\{ \sigma_{\bs{s}_1}^{\otimes M}, \sigma_{\bs{s}_2}^{\otimes M}, \ldots, \sigma_{\bs{s}_N}^{\otimes M}\}$ are irradiated. More precisely, we can define a probe-domain distribution as
\begin{gather}
\mc{S} \defeq \{\bs{s}_1, \bs{s}_2,\ldots,\bs{s}_N\} = \bigcup_{j=1}^{N} \{ \bs{s}_j \}, \label{eq:ProbeDistrib}\\
\text{} \exists\> j \text{ such that } i \in \bs{s}_j, \forall i \in \{1,2,\ldots, m\}.   \label{eq:allcontained}
\end{gather}
In Eq.~(\ref{eq:allcontained}), we demand that every channel index $1,\ldots, m$ is accounted for in at least one subset $\bs{s} \in \mc{S}$, so that no channels are left un-probed. Using $\mc{S}$ we can define a global probe state irradiated over a channel pattern, constructed as the tensor product of all the local sub-states.

It is not immediately clear how one should design this probe-domain distribution. However, the most intuitive way to construct $\mc{S}$ is to devise a distribution such that each channel is only associated with a single probe-domain. A probe-domain distribution \textit{disjoint} if it satisfies this property. Suppose one constructs a $N$-partite probe-domain distribution that is disjoint, $\mc{S}_{\text{d}}$. 
Formally, we can define this as,
\begin{gather}
\mc{S}_{\text{d}} \defeq \bigcup_{j=1}^{N} \{ \bs{s}_j\},~\text{such that}~\bs{s}_j \cap \bs{s}_k = \varnothing,~\forall j,k, \label{eq:DisjDef}
\end{gather} 
where disjointedness is demanded on the RHS of this equation, such that no two probe-domains $\bs{s}_j$ and $\bs{s}_k $ are permitted to share the same channel label, for all $j,k$. Again, we demand that all channels $1,\ldots,m$ are accounted for in this distribution, as in Eq.~(\ref{eq:allcontained}). 
We may then choose an $N$-element set of multipartite probe states in accordance with this disjoint structure $\{\sigma_{\bs{s}_j} \}_{j=1}^N$, where each $\sigma_{\bs{s}_j}$ can be unique. Assuming $M$-copies of each sub-state, we can define a global probe state
\begin{equation}
\sigma_{\mc{S}_{\text{d}} }^{\otimes M} = \sigma_{\bs{s}_1}^{\otimes M} \otimes \ldots \otimes \sigma_{\bs{s}_N}^{\otimes M} = \bigotimes_{j=1}^{N} \sigma_{\bs{s}_j}^{\otimes M} .
\end{equation}
In this way, each channel in the pattern is probed exactly $M$ times per total round of discrimination. Furthermore, since all probe-domains are disjoint, there are no overlaps between any multipartite states; each channel in the pattern is always probed within the same probe-domain and within the same collection of channels.\par

From an operational point of view, the disjointedness of $\mc{S}_{\text{d}}$ and lack of probe-domain overlaps means that each sub-state $\sigma_{\bs{s}_j}^{\otimes M}$ can interact simultaneously with the multi-channel.
As such, each probe state can be considered to be static (or \textit{fixed}) over a sub-region of the channel pattern throughout the entire discrimination protocol. For this reason, we describe an unassisted protocol using a disjoint probe-domain distribution as a \textit{fixed block protocol}, $\mc{B}_{\text{fix}}^{\text{u}}$ (see Fig.~\ref{fig:DisjMap}(a) for an example). \par

Fixed block protocols are very intuitive thanks to their simple, static format. Indeed, classical block protocols can inherently be considered to be fixed protocols, where separable collections of coherent states are irradiated upon a channel pattern. Using our previous formalism and considering $m$-length channel patterns, one may define a trivial probe-domain distribution $\mc{S}_{\text{d}}  = \{ \{1\}, \{2\}, \ldots, \{m\}\}$ and a corresponding set of single-mode coherent states $\{ \alpha_j \}_{j=1}^m$ which produces the global state ${\sigma_{\mc{S}_{\text{d}} }^{\otimes M} = \bigotimes_{j=1}^m \alpha_j^{\otimes M}}$. Larger probe-domains invite the potential for entangled probe states over fixed probe-domains, and can provide an easy route for potential quantum enhancements in many settings. In general, the performance of fixed block protocols can always be assessed through the average error probability by substituting $\sigma_{\mc{S}_{\text{d}}}$ into Eq.~(\ref{eq:p_err_block}).

\subsection*{Dynamic Unassisted Block Protocols}

Interestingly, we need not restrict ourselves to probe-domain distributions which are disjoint. Departing the rigidity of disjoint probe-domain distributions offers a fascinating route for quantum-enhanced, unassisted protocols. While this path is less intuitive, it unveils a rich and flexible class of discrimination protocols with rewarding features.

Consider now a non-disjoint, $N$-partite probe-domain distribution ${\mc{S}_{\text{nd}} = \bigcup_{j=1}^{N}\{ \bs{s}_j} \}$, meaning that probe-domains are free to overlap and share similar channel labels, i.e.~the overlap of two probe-domains is no longer the empty set $\bs{s}_j \cap \bs{s}_k \neq \varnothing$. This renders a much larger, and more general space of possible distributions. A global quantum probe state $\sigma_{\mc{S}_{\text{nd}}}$ associated with such a distribution is again found as the tensor product of all local sub-states, however its interpretation is much less obvious. We begin by describing the physical interpretation of a non-disjoint probe-domain distribution within a discrimination protocol.

 \begin{figure}
 \hspace{-0.7cm} (a) Disjoint $\mathcal{S}_{\text{d}}$,  \hspace{1.5cm} (b) Non-Disjoint $\mathcal{S}_{\text{nd}}$. \\
\hspace{0cm}\\
\includegraphics[width=\linewidth]{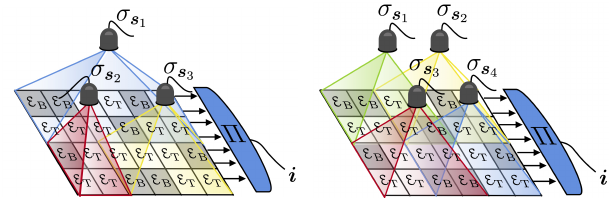}
\caption{\textbf{Unassisted Block Protocols:} (a) Disjoint vs.~(b) non-disjoint multipartite probe-domain distributions assuming the use of single-copy probe states, $M=1$. In (a) there are clearly no overlapping probe-domains, and it therefore generates a fixed block protocol. Contrarily, the overlapping probe-domains in (b) gives rise to a dynamic block protocol.}
\label{fig:DisjMap}
\end{figure}

Any non-disjoint discrete distribution $\mc{S}_{\text{nd}}$ can be decomposed into a sequence of $r$ disjoint distributions,
\begin{equation}
\mc{S}_{\text{nd}} = \bigcup_{k=1}^{r}  \mc{S}_{\text{d}}^k  = \bigcup_{k=1}^{r} \bigcup_{\bs{s} \in \mc{S}_{\text{d}}^k} \{ \bs{s} \} , \label{eq:NDisjDef}
\end{equation}
where $\mc{S}_{\text{d}}^k$ is a disjoint sub-collection of probe-domains in accordance with Eq.~(\ref{eq:DisjDef}). In this case, each $\mc{S}_{\text{d}}^k$ need not contain all the channel labels, but all $m$ channels must be accounted for in the global distribution $\mc{S}_{\text{nd}}$. This allows us to rewrite the global, single-copy probe state in a more meaningful way 
\begin{equation}
\sigma_{\mc{S}_{\text{nd}}} = \bigotimes_{k=1}^r  \sigma_{\mc{S}_{\text{d}}^k} = \bigotimes_{k=1}^r \bigg[ \bigotimes_{\bs{s}\in \mc{S}_{\text{d}}^k} \sigma_{\bs{s}}\bigg].
\end{equation}
That is, it is the tensor product of $r$ disjointly distributed multipartite input states. \par
\begin{figure*}
 \hspace{2cm}(a) Dynamic Block Protocol, \hspace{5cm} (b) Fixed Representation. \\
 \hspace{1cm}\\
\includegraphics[width=\linewidth]{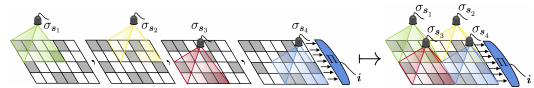}\\
 \hspace{3cm} Original channel pattern $\bs{i}$, \hspace{6cm} Modified pattern $\bs{\nu_i}$.
\caption{ \textbf{Dynamic to Fixed Block Protocol Transformation:} (a) Non-disjoint probe-domain distributions can be decomposed into multiple rounds of disjoint pattern interaction, generating a dynamic discrimination protocol. 
This dynamic protocol can be equivalently represented by a fixed block protocol on a modified image space, as shown in (b). The original $(6\times 4)$-channel pattern $\bs{i}$ is transformed into a $(8\times 4)$-pattern $\bs{\nu_i}$ which has been appropriately modified in accordance with the non-disjoint probe-domain distribution $\mc{S}_{\text{nd}}$ using Eq.~(\ref{eq:PattTransf}). Here we have assumed the use of single-copy probe states, $M=1$.}
\label{fig:Dynamic}
\end{figure*}

 Therefore, the utilisation of a non-disjoint probe-domain distribution corresponds to a block protocol with $r$ rounds of disjoint pattern interaction. At each round, the user interacts with the channel pattern by irradiating unassisted multipartite states, and over the course of $r$ rounds the probe-domain distribution  ``moves" around the channel pattern. For this reason, it can be intuitively called a \textit{dynamic block protocol}, $\mc{B}_{\text{dy}}^{\text{u}}$. Fig.~\ref{fig:DisjMap}(b) depicts an $m=4\times 6 = 24$ channel pattern which is being non-disjointly probed. The dynamic ``movement" of probe-domains throughout its $r=4$ rounds of disjoint pattern interaction is visualised in Fig.~\ref{fig:Dynamic}(a).
\par
 
 The number of disjoint rounds $r$ required to construct a dynamic protocol depends on the number of overlaps that occur within the decomposition in Eq.~(\ref{eq:NDisjDef}). An overlap simply refers to an instance of a channel label that is contained in more than one probe-domain. We can define the number of overlaps $m_{\text{ov}}$ as the total number of additional channel labels contained in the non-disjoint distribution
 \begin{equation}
 m_{\text{ov}} \defeq \big[\sum_{\bs{s}\in\mc{S}_{\text{nd}}}\hspace{-0.0mm} |\bs{s}| \big] - m.
 \end{equation} 
 If there are many probe-domain overlaps then $r$ may be very large; if there are no overlaps, then $r = 1$ and we return to a fixed protocol. 

In order to fairly compare dynamic and fixed block protocols, one must also be careful when distributing the number of probe copies $M$; a dynamic protocol with $r$ rounds of disjoint pattern interaction and $M$-copy input states will clearly use more than $M$ total probe modes. It is useful to define a resource metric known as the \textit{average channel use}, \begin{equation}
\bar{M} \defeq \frac{m + {m_{\text{ov}}}}{m} M,
\end{equation}
which describes average number of probe copies applied per channel within a dynamic block protocol. When comparing the performance of fixed/dynamic block protocols, we must ensure they have the same average channel use.

\subsection*{Dynamic/Fixed Block Protocol Transformation}

Consider a dynamic block protocol which follows a non-disjoint probe-domain distribution $\mc{S}_{\text{nd}}$. Now, any channel $\mc{E}_{i_j}$ within the global pattern $\mc{E}_{\bs{i}}$ may be probed as part of several different multipartite domains. This more general scenario requires a mathematical model that allows us to quantitatively investigate the performance of dynamic protocols.

To achieve this, we find a simple relationship between dynamic and fixed block protocols, corresponding to an appropriate transformation on a channel pattern image space, $\mc{U}$. When two probe-domains overlap, the overlapping channels are probed twice, but by independent probe states. Therefore, we attribute a unique Hilbert space to each independent probe mode and channel in each disjoint round throughout the protocol, whilst retaining the characteristics of the original channels. This can be done by considering a modified channel pattern which has been concatenated with \textit{copies} of the channels that are overlapped. \par 

Fig.~\ref{fig:Dynamic} depicts how this pattern modification takes place. Given that $\mc{S}_{\text{nd}}$ contains $m_{\text{ov}}$ overlapping channels,
an originally $m$-length channel pattern $\bs{i}$ can be mapped to a $(m+m_{\text{ov}})$-length pattern, where the additional copies of overlapping channels are concatenated with the multi-channel. These copy channels directly obey the behaviour of their originals. In this way, a dynamic protocol over $m$-length channel patterns can be equivalently studied as a fixed block protocol over an appropriately modified $(m+m_{\text{ov}})$-length channel pattern space. \par

Let us more precisely express this transformation. A $\mc{S}_{\text{nd}}$ dynamic protocol invokes the following transformation on a generic $m$-length channel pattern $\bs{i}$ into an extended channel pattern $\bs{\nu_i}$. Since $\bs{\nu}_{\bs{i}}$ contains repeated elements it is formally treated as a \textit{multi-set}, rather than a traditional set \cite{MultiSets}. Then the transformation can be explicitly written as
\begin{equation}
\bs{i} = \{i_{1}, i_2, \ldots, i_m \} \mapsto \bs{\nu_i} \defeq {\biguplus_{\bs{s}\in \mc{S}_{\text{nd}}}} \{ i_{k} \}_{k \in \bs{s}}. \label{eq:PattTransf}
\end{equation} 
where $\uplus$ is the multi-set union operator which concatenates each subset of channel labels, e.g.~if we consider $m=3$ length channel patterns and a probe-domain distribution $\mc{S}_{\text{nd}} = \{\{1,2\}, \{2,3\}\}$ then modified channel patterns take the form $\bs{\nu_i} = \{i_1,i_2\} \uplus \{i_2,i_3\} = \{ i_1,i_2,i_2,i_3\}$. From a channel perspective, this transformation can be equivalently portrayed as
\begin{equation}
\mc{E}_{\bs{i}} = \mc{E}_{i_{1}} \otimes\ldots\otimes \mc{E}_{i_{m}} \mapsto \mc{E}_{\bs{\nu_i}}\defeq \bigotimes_{\bs{s}\in \mc{S}_{\text{nd}}} \bigotimes_{k \in \bs{s}} \mc{E}_{i_{k}}.
\end{equation}
By iterating this concatenation process over all patterns in an image space $\{ \bs{\nu}_{\bs{i}} \}_{\bs{i} \in \mc{U}}$, one can easily convert a dynamic protocol into a fixed representation. Furthermore, it is expedient to write the global output states of these protocols in this format, such that
\begin{equation}
\sigma_{\bs{i}} \mapsto \sigma_{\bs{\nu_i}} \defeq \mc{E}_{\bs{\nu_i}} \left( \sigma_{\mc{S}_{\text{nd}}} \right).
\end{equation}

This transformation greatly simplifies the complication of overlapping probe-domains, and allows for an investigation of error probabilities. By abstracting our set of discriminatory POVMs to the modified image space $\{{\Pi}_{\bs{\nu}_{\bs{i}}}\}_{\bs{i}\in \mc{U}}$, and using an $M$-copy global probe state, then the average error probability of classification can be succinctly given by
\begin{equation}
p_{\text{err}}(\mc{B}_{\text{dy}}^{\text{u}}) = \sum_{\bs{i} \neq \bs{i}^{\prime} \in \mc{U}} \pi_{\bs{i}} \text{Tr}\left[ {\Pi}_{\bs{\nu}_{\bs{i}^{\prime}}} \sigma_{\bs{\nu_i}}^{\otimes M}\right] .
\end{equation}
Without explicit knowledge of these POVMs, we can simply utilise the fidelity bounds from Eqs.~(\ref{eq:LB}) and (\ref{eq:UB}). These fidelity-based bounds can be readily computed thanks to this fixed protocol transformation, by iterating over all unequal channel patterns in the modified image space (see Methods for more details).\\

\subsection*{Correspondence with Error Correction}

This dynamic to fixed block protocol mapping identifies a fascinating feature. In essence, a dynamic protocol invokes an \textit{encoding} of quantum channel patterns, wherein $m$-length patterns from some image space $\bs{i}\in\mc{U}$ are encoded into an extended counterpart $ \{ \bs{\nu_i} \}_{\bs{i}\in \mc{U}}$. This modified image space is a function of the non-disjoint probe-domain distribution $\mc{S}_{\text{nd}}$. Thus, we make the crucial observation: Using entangled probe states, one can design a dynamic block protocol which encodes a quantum image space into a more distinguishable form.\par

Consider a single channel $\mc{E}_{i_{\text{ov}}}$  within a larger-channel pattern, which happens to fall within the domain of two entangled sub-states of a global probe, ${\sigma_{\mc{S}_{\text{nd}}} = \sigma_{\bs{s}}\otimes \sigma_{\bs{s}^{\prime}}}$. Because of this, the probe states must be applied at different disjoint rounds in a dynamic protocol. In one round, the probe state $\sigma_{\bs{s}}$ is being used to determine the classification of all the channels ${\mc{E}_{\bs{i}[\bs{s}]} \defeq \{\mc{E}_{i_{k}} \}_{k\in \bs{s}}}$. In another round, the probe state $\sigma_{\bs{s}^{\prime}}$ is being used to classify the channels in the region ${\mc{E}_{\bs{i}[\bs{s}^{\prime}]} = \{\mc{E}_{i_{k}} \}_{k\in \bs{s}^{\prime}}}$. Because these probe sub-states are entangled over their domains, then the distinguishability of their output states $ \mc{E}_{\bs{i}[\bs{s}]} (\sigma_{\bs{s}})$ and $ \mc{E}_{\bs{i}[\bs{s}^{\prime}]} (\sigma_{\bs{s}^{\prime}})$ are correlated with the precise collection of quantum channels in each region. Dependent upon the size of entangled probe domains and the physical setting of discrimination, some collections of channels are easier to discriminate than others.\par

We arrive at the key insight:
Since $\mc{E}_{i_{\text{ov}}}$ is contained in both probe-domains, we are able to gather two potentially unique \textit{opinions} on its classification; one from the perspective of $\sigma_{\bs{s}}$ in the pattern region $\bs{s}$, and another from $\sigma_{\bs{s}^{\prime}}$ in its region $\bs{s}^{\prime}$. On their own, these states may not be very effective at discriminating the channel $\mc{E}_{i_{\text{ov}}}$, i.e.~one of the output states may not be very distinguishable from from other potential output states in that region. But by probing $\mc{E}_{i_{\text{ov}}}$ in conjunction with two different probing domains, it is more likely that \textit{at least one} of the sub-regions will be a more distinguishable collection of channels; thus providing a greater chance of correct classification.\par

In this way, dynamic block protocols implicitly possess a form of \textit{error-correcting} behaviour. By varying the spatial probe-domain distributions throughout the protocol, channels are probed from various perspectives, correlated with different sub-regions of the channel pattern. Poorly distinguishable channels in one sub-region may be significantly more distinguishable when probed within a different sub-region. Over the course of $r$ disjoint rounds of pattern interaction, each entangled multipartite sub-state can help to correct errors that would arise if only fixed probe-domains were used. Exploiting this behaviour, dynamic protocols can indeed encode channel patterns into more easily discriminated image spaces.

This is a remarkable property of dynamic block protocols, one that depends strongly on the choice of entangled quantum probes and the quantum channel patterns. Explicit examples of this behaviour and physical/mathematical intuition are elucidated in the Methods section.

\begin{figure*}[t!]
\includegraphics[width=0.9\linewidth]{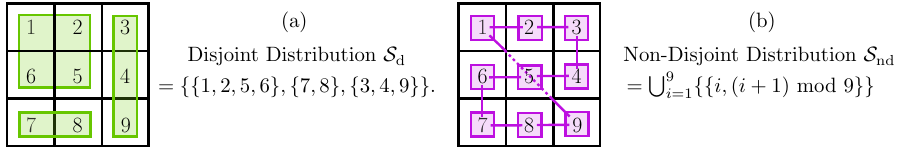}
\caption{\textbf{Diagrammatic Protocol Representation}: (a) Disjoint and (b) non-disjoint distributions of multipartite probe-domains. The example in (b) is in fact the nearest-neighbour configuration described in Eq.~(\ref{eq:NNdistr}).  }
\label{fig:expl_diags}
\end{figure*}

\subsection*{Designing Unassisted Block Protocols}
Given an $m$-length channel pattern discrimination problem, there are clearly an enormous number of ways in which one can design a (generally non-disjoint) probe-domain distribution. Let us provide a diagrammatic approach to constructing these protocols.

An $m$-length channel pattern $\mc{E}_{\bs{i}}$ can be represented by an $m$-pixel grid. Each pixel is used to represent an individual channel $\mc{E}_{\bs{i}_j}$, and the grid can adopt any preferred the height, width and shape.
In order to create a tidy language that allows one to convey probe-domains which contain both local and non-local channels, we provide two ways to portray a probe-domain. Firstly, a probe-domain can be indicated by means of a coloured box which covers a number of local channels. The size and position of the coloured-probing domains indicate the regions of a channel pattern that are irradiated by a multipartite input state. This is particularly useful for illustrating fixed block protocols with local probe-domains, as shown in Fig.~\ref{fig:expl_diags}(a), which can be intuitively interpreted. 

It is also useful to possess a convention for probe-domains which is more convenient with non-local channels, or when are a number of overlapping domains in close proximity. 
Hence, we can equivalently illustrate a probe-domain via connective lines between coloured single-pixel boxes. A probe-domain is indicated by means of a continuous (unbroken) connecting line between a number of pixels. Dashed connective lines through channel boxes indicate a lack of entanglement, used to bypass certain channels while illustrating non-local domains. A clear example of this is shown in Fig.~\ref{fig:expl_diags}(b). Here, we describe a distribution of probe-domains where each domain is of size $|\bs{s}| =2$, i.e.~$\mc{S}_{\text{nd}} = \{\{1,2\},\{2,3\},\ldots,\{8,9\},\{9,1\}\}$. A dashed connective line is used to connect the non-local channel labels $1$ and $9$ so that $5$ is not included in the probe-domain.

\subsection*{Bosonic Gaussian Channel Patterns}

We wish to corroborate the construction of these classes of unassisted discrimination protocols and demonstrate their efficacy. To do so, we will focus on the discrimination of bosonic Gaussian Phase Insensitive (GPI) channels. This is a family of very important channels within CV quantum communications, sensing and computation \cite{GaussRev}, and can be used to model a vast array of physically significant scenarios. Crucially, a GPI channel maintains the Gaussianity of its input state. Hence, the transformation of a Gaussian state (with zero first moments) through under the action of a GPI channel can be fully characterised using its covariance matrix $V$, assuming zero first moments (see Methods for the explicit transformations). The overall quantum channel can be denoted as $\mc{E}_{\tau,\nu}$ and is defined with respect to a transmissivity parameter ${0 \leq \tau \leq 1}$ describing attenuation/amplification properties and an induced noise parameter $\nu \geq 0$. When $\tau = 1$ and $\nu = 0$ we regain the identity channel. \par

Binary GPI channel patterns then consist of a sequence of $m$ GPI channels with unique target/background transmissivities $\tau_B, \tau_T$ and noise properties $\nu_B, \nu_T$. Generally, we may write the channel pattern
\begin{equation}
\mc{E}_{\bs{i}} = \mc{E}_{\tau_{i_1}, \nu_{i_1}} \otimes \ldots \otimes \mc{E}_{\tau_{i_m}, \nu_{i_m}} = \bigotimes_{j=1}^m \mc{E}_{\tau_{i_j}, \nu_{i_j}}.
\end{equation}
Let us identify some essential GPI channels: Setting $\tau = \eta$, such that $0 < \eta < 1$ and ${\nu = ({1- \eta})/{2}}$ then we have the single parameter bosonic pure-loss channel $\mc{E}_{\eta}$. This describes the interaction of bosonic mode with a zero-temperature bath. This is an essential channel model for the description of optical fibres, and short-range optical target detection known as quantum reading. The multi-hypothesis setting of discrimination pure-loss channel patterns has also be equated to the task of optical imaging, pattern recognition and classical data-readout from optical memories \cite{PirBD, PatternRecog}. Hence, the discrimination of bosonic loss poses a key problem setting for our work.\par

Alternatively, we may study thermal-loss channels $\mc{E}_{\tau,\nu}$ such that the transmissivity satisfies $0 < \tau < 1$, or thermal-amplifier channels where $\tau > 1$. In both cases, the induced thermal noise is connected to the number of thermal photons in the channel environment $N_{\text{env}}$, such that $\nu = (N_{\text{env}}+\frac{1}{2})|1-\tau|$. In the idealised absence of loss, we have a Gaussian additive-noise channel $\mc{E}_{\nu}$, where the transmissivity satisfies $\tau = 1$ but we have non-zero noise $\nu > 0$. The discrimination of thermal multi-channels is known as environment localisation, and  has been used to model fascinating scenarios within target detection and thermal pattern recognition \cite{OptEnvLoc, ThermalPatt}. In this work, we focus on the discrimination of additive-noise binary channel patterns, since the performance of this task will always be an upper bound for multi-channels with non-trivial transmissivity.

\subsection*{Unassisted Bosonic Quantum Pattern Recognition}
In order to devise fixed/dynamic unassisted block protocols for the discrimination of GPI channel patterns, we must specify a class of multipartite probe state. Here, we make use of the Gaussian analogue of the entangled GHZ state known as a CV-GHZ state which is designed as the extension of a TMSV state to many modes. 
Consider a probe-domain $\bs{s}$ which describes a collection of $|\bs{s}|$ channels over which an input probe state is irradiated. A CV-GHZ state defined over this probe-domain is a $|\bs{s}|$-mode, fully symmetric state denoted by $\Phi_{\bs{s}}^{\mu}$. It can be completely characterised by its covariance matrix (CM) with zero first moments \cite{multimodeGHZ},
\begin{equation}
\Phi_{\bs{s}}^\mu \mapsto V_{\bs{s}}^{\mu} \defeq \begin{pmatrix}
\mu I & \Gamma & \ldots & \Gamma \\
\Gamma  & \mu I & \ldots & \Gamma \\
\vdots &\ddots  & \ddots & \vdots\\
\Gamma & \Gamma & \ldots & \mu I
\end{pmatrix},
\begin{tabular}{ l } 
$\mu \defeq N_S + \frac{1}{2}$,  \\
\\
$\Gamma \defeq \text{diag}(c_1,c_2)$.
\end{tabular}
\label{eq:CVGHZ}
\end{equation}
Here, $\mu$ denotes the energy (squeezing) of the state for shot noise ${1}/{2}$ and mean photon number (or signal energy) $N_S$, and $I$ denotes the $2\times 2$ identity matrix. Therefore, $V_{\bs{s}}^{\mu}$ is a $2|\bs{s}|\times 2|\bs{s}|$ real matrix which is fully symmetric. In order to capture maximal correlations at finite squeezing, we set the correlation parameters 
\begin{equation}
c_1 = -c_2 = c_{\text{max}} \defeq {\sqrt{\mu^2 - 1/4}}/({|\bs{s}|-1}). \label{eq:Correlations}
\end{equation}
See the Methods section for more details on this state. Hence, we may construct unassisted, global quantum probe states from CV-GHZ sub-states. Given an arbitrary $N$-partite probe-domain distribution $\mc{S}$, and assuming that all sub-states are of the same energy $\mu$, the global $M$-copy input state is given by
\begin{equation}
\sigma_{\mc{S}}^{ \otimes M} = \Phi_{\mc{S}}^{\mu \otimes M} = \bigotimes_{j=1}^{N} \Phi_{\bs{s}_j}^{\mu \otimes M}.\label{eq:GenGHZState}
\end{equation}

As seen in Eq.~(\ref{eq:Correlations}), the magnitude of the correlations held by CV-GHZ states $c_{\max}$ has a reciprocal dependence on the number of modes $m$ in the state. This implies that the quantum correlations become ``thinner" as the number of modes increase, demanding more energy in order to maintain a high degree of entanglement. It is therefore beneficial to consider probe-domain distributions of shorter range CV-GHZ states in order to make the best use of the enhanced distinguishability that entanglement can provide. Motivated by this, we can design specific probe-domain distributions that exclusively use unassisted TMSV entangled states, rather than wider-spread CV-GHZ states with weaker quantum correlations, i.e.~the probe-domain of each sub-state is simply $|\bs{s}_j| = 2$, $\forall j$. By employing TMSV states in conjunction with dynamic block protocols, we can combine the enhanced distinguishability of entangled input states with the implicit error-correcting behaviour available through dynamic probing.\par

To systematically access both of these features, we introduce a \textit{nearest-neighbour} probe-domain distribution. This defines a non-disjoint probe-domain distribution which probes neighbouring channels using two-mode probe states (defining neighbouring channels on a closed 1-dimensional lattice). The non-disjoint partition set takes the form,
\begin{equation}
\mathcal{S}_{\text{nd}} =  \bigcup_{i=1}^{m} \{ \{i,(i+1) ~\text{mod}~ m\}\} . \label{eq:NNdistr}
\end{equation}
where $\text{mod}$ denotes the modulo operation. For example, if $m=4$, the probe-domain distribution is simply ${\mathcal{S}_{\text{nd}} = \{ \{1,2\}, \{2,3\}, \{3,4\}, \{4,1\}\}}$. In this way, each channel in the global pattern is probed exactly twice per single round of a dynamic block protocol (the average channel use is simply $\bar{M} = 2M$). Diagrammatically, this distribution is illustrated in Fig.~\ref{fig:expl_diags}(b).
The nearest-neighbour protocol is conveniently designed, as it allows us to develop non-disjoint probing structures in a consistent way, and can be applied to channel patterns of any size (for more detailed arguments and motivations surrounding this protocol, see the Methods section).

\begin{figure*}[t!]
\includegraphics[width=\linewidth]{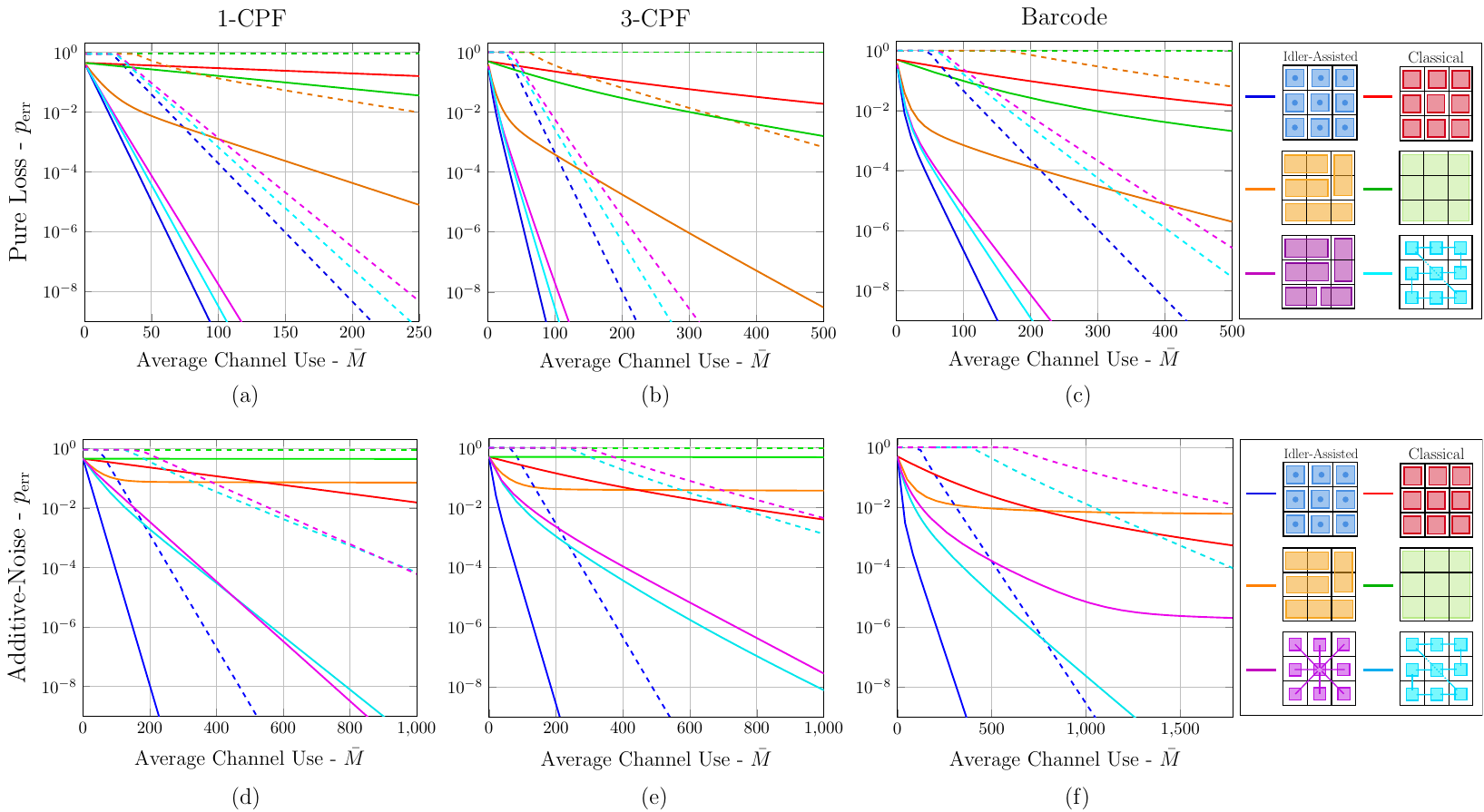}
\caption{\textbf{Discrimination of Bosonic Gaussian Channel Patterns:} Classification error bounds for CPF/Pattern recognition of $m=9$ channel patterns of (a)-(c) Pure Loss Channels with parameters $ \eta_T, \eta_B = 0.97, 0.99 $ and (d)-(f) Additive Noise Channels with parameters $\nu_T, \nu_B = 0.01,0.02$, using probe states of mean photon energy $N_S = 20$ and variable structures based on CV-GHZ states (and optimal classical states). All solid lines are lower bounds and all dashed lines are upper bounds, based on Eqs.~(\ref{eq:LB}) and (\ref{eq:UB}) respectively. All input state structures are defined diagrammatically in the respective legends.}
\label{fig:PattFids}
\end{figure*}

\begin{figure*}[t!]
\includegraphics[width=\linewidth]{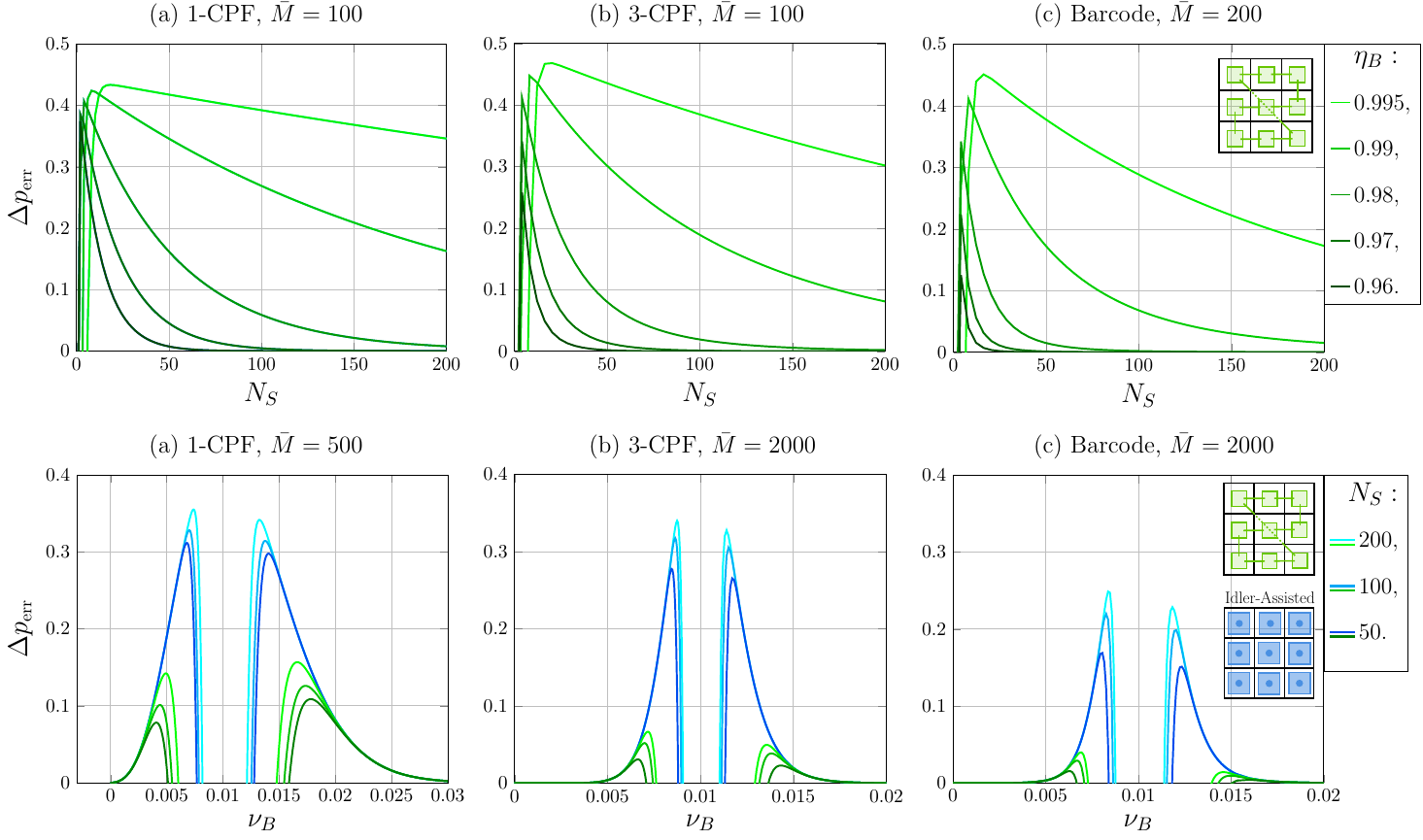}
\caption{\textbf{Guaranteed Quantum Advantage:} as per Eq.~(\ref{eq:GAdv}), for 9-pixel (a),(d) CPF,  (b),(e) 3-CPF, and (c),(f) full image space (barcode) discrimination using the nearest-neighbour dynamic protocol compared with full idler-assistance. In panels (a)-(c) the $m=9$ pure-loss channels are considered, with target pixel transmissivity $\eta_T = 1$ and various background transmissivities $\eta_B$, plotted against signal energy $N_S$. Here the difference in advantage with the idler-assisted protocol is too small to be plotted. Panels (d)-(f) study additive noise $m=9$ channel patterns for target noise $\nu_T = 0.01$ and various signal energies, plotted against background noise $\nu_B$.  }
\label{fig:EnergyPlots}
\end{figure*}

\subsection*{Numerical Results}

In this section, we collect numerical results to benchmark the performance of both fixed and dynamic unassisted block protocols for the discrimination of bosonic pure-loss channel patterns (quantum reading) and Gaussian additive-noise channel patterns (environment localisation). We investigate a number of pattern recognition scenarios: CPF, $k$-CPF, and arbitrary binary pattern classification (or barcode decoding). In each setting, we consider the worst-case discrimination scenario such that all patterns within an image space occur with a uniform probability, i.e.~we consider the pattern probability distribution 
\begin{equation}
\pi_{\bs{i}} = {|\mc{U}|}^{-1}, ~\forall {\bs{i}\in\mc{U}}.
\end{equation} \par

In all cases we employ unassisted CV-GHZ states in accordance with various disjoint/non-disjoint probe-domain distributions. The average error probability associated with these protocols can be accurately upper and lower bounded using the fidelity bounds in Eqs.~(\ref{eq:LB}) and (\ref{eq:UB}) for which a variety of numerical and analytical techniques can be used for arbitrary multipartite distributions (see Methods for details on the numerical computations). \par

These unassisted protocols can be compared to the best known classical and quantum assisted protocols in order to critically benchmark their efficacy (details can be found in Methods). A sufficient condition for quantum advantage occurs when the upper bound for the average error probability associated with a quantum enhanced protocol $p_{\text{err}}^{\text{q,U}}$ is less than a lower bound on the error probability associated with an optimal classical protocol $p_{\text{err}}^{\text{cl,L}}$. Hence, we may qualify guaranteed quantum advantage when
\begin{equation}
\Delta p_{\text{err}} = p_{\text{err}}^{\text{cl,L}} - p_{\text{err}}^{\text{q,U}} \geq 0. \label{eq:GAdv}
\end{equation}
We use this quantity $\Delta p_{\text{err}}$ to identify when an unassisted quantum protocol can certifiably obtain quantum advantage over all classical protocols. 

\subsection*{Discrimination of Bosonic Loss}
This can be used to describe a basic imaging setting, in which pixels are described by pure-loss channels of different transmissivity/reflectivity $\eta_j$ for $j \in \{B,T\}$. As explored in \cite{PatternRecog}, Banchi et al.~showed that major quantum advantage can be obtained using an idler-assisted approach. This advantage is particularly useful in a low energy regime, where the number of probe copies required to achieve high precision is dramatically reduced. Here we report that quantum advantage can be similarly guaranteed using a range of unassisted protocols. Moreover, it is possible to achieve unassisted performances comparable with that of full idler-assistance via dynamic block protocols. \par

Figs.~\ref{fig:PattFids} (a)-(c) depicts error upper and lower bounds for the multi-channel discrimination of bosonic loss (upper bounds are plotted as dashed lines, lower bounds are solid).
We consider $m=9$ binary channel patterns such that background channel possess transmissivity $\eta_B= 0.99$, while target channels possess $\eta_T = 0.97$. In each panel (a)-(c) we consider a different image space: CPF, $(k=3)$-CPF and barcode pattern recognition respectively. Within each setting, we construct fixed and dynamic unassisted block protocols using CV-GHZ sub-states, each with mean photon energy $N_S = 20$. The precise probe-domain distributions are identified diagrammatically in the legend.\par

Fig.~\ref{fig:PattFids}(a) shows results for CPF. While one can eventually confirm quantum advantage using a block protocol with a single $m=9$ mode CV-GHZ state (as studied in Ref.~\cite{Jason_IdlerFree}), this is only certifiably advantageous using a very large average channel use, $\bar{M} \approx 3000$, compared to the idler-assisted protocol $\bar{M} \approx 30$. Furthermore, for the larger image spaces it quickly becomes too costly to achieve guaranteed quantum advantage, such as for 3-CPF and barcode discrimination. Instead, one may use a dynamic protocol to achieve error rates on par with the idler-assisted performance. Using the nearest-neighbour dynamic protocol as per the probe-domain distribution in Eq.~(\ref{eq:NNdistr}), one may readily obtain guaranteed quantum advantage regardless of the image space. This dynamic protocol not only outperforms the optimal classical protocol, but also quickly provides guaranteed advantage over the best fixed unassisted block protocols also, achieving performance on par with idler-assistance. \par

Fig.~\ref{fig:EnergyPlots}(a)-(c) displays the minimum guaranteed quantum advantage $\Delta p_{\text{err}}$ associated with the use of nearest-neighbour dynamic protocols. Here we plot the difference between the quantum upper bound and the optimal classical lower bound for $m=9$ channel pattern discrimination. This is carried out for $\eta_T=1$, $\bar{M}=100$, a variety of signal energies $N_S$, and background transmissivities $\eta_B$. The difference in advantage achieved by the idler-assisted protocol and the nearest-neighbour dynamic protocol is too small to be displayed; emphasising that we can not only achieve quantum advantage without idlers, but effectively match the performance of idler-assistance. \par

\subsection*{Environment Localisation}
We now consider environment localisation. Here the task is to classify channel patterns in which each channel possesses background or target noise properties, $\nu_B/\nu_T$. Note that we focus on additive noise channels as an idealised scenario for thermal-loss/amplifier channels since the inclusion of loss $\tau \neq 1$ will only degrade the performance of our unassisted protocols. It has recently been proven that the ultimate error bounds for this pattern recognition setting are non-adaptively achieved by idler-assisted TMSV states \cite{OptEnvLoc,ThermalPatt}. \par

In Figs.~\ref{fig:PattFids}(d)-(c) we report the performance of a number of different fixed/dynamic unassisted protocols for the task of environment localisation. Again, we consider $m=9$ length channel patterns for a trio of image spaces, CPF, $(k=3)$-CPF and barcode pattern recognition. Each channel is characterised as an additive noise channels with $\nu_B = 0.02$ or $\nu_T = 0.01$, and our probe states again have mean photon number $N_S=20$.
It is immediately clear that unassisted, fixed block protocols in this setting are ineffective, as shown by the very poor lower bounds in these results. Without idlers, the output distinguishability of disjointly distributed probe-states is extremely poor, and degrades further with increasing probe-domain size.\par

Yet, performance can be redeemed via dynamic protocols. By overlapping entangled probe-domains over channel patterns, we increase the opportunity of interacting with distinguishable channel regions. Indeed, the use of the nearest-neighbour dynamic protocol allows for guaranteed quantum advantage to be obtained in a number of discrimination settings where fixed protocols are unable to even match the classical performance (see Methods for more nuanced insight to this result). Interestingly,  alternative non-disjoint probe-domain distributions can be seen to achieve quantum advantage also, in some cases outperforming the nearest-neighbour protocol as shown in Fig.~\ref{fig:PattFids}(d). The question of identifying optimal dynamic protocols is thus highly non-trivial and very interesting. \par

Finally, Figs.~\ref{fig:EnergyPlots}(d)-(f) compare the guaranteed quantum advantage $\Delta p_{\text{err}}$ associated with idler assisted protocols with that of the nearest-neighbour dynamic protocol in this discrimination setting, for $\nu_T = 0.01$ and a variety of resource/environmental parameters. The most significant guaranteed advantage is observed for 1-CPF, as shown in both Fig.~\ref{fig:PattFids}(d) and Fig.~\ref{fig:EnergyPlots}(d). While it is clear that unassisted protocols are more sensitive to noisy, thermal environments, quantum advantage is still achievable without the use of idlers. These results emphasise the achievability of quantum-enhanced, idler-free protocols for short-range environment localisation tasks.

\section*{Discussion\label{sec:Conclusions}}

We have formalised the construction of unassisted, quantum enhanced discrimination protocols in a multi-channel setting, using multipartite quantum states. We identified two distinct classes of block protocols, fixed and dynamic, which differ in how they distribute multipartite entanglement across channel patterns. The operational interpretations of these protocols were discussed, along with their relationship with one another.  Furthermore, we formulated a logical correspondence between dynamic protocols and error correction; variable probe-domains throughout discrimination help to correct errors that fixed probe-domains cannot. 

In order to explicitly study the efficacy of these protocols, we designed unassisted protocols for the discrimination of bosonic Gaussian channel patterns. These protocols were based based on the use of entangled, multi-mode CV-GHZ states. Through analytical and numerical investigation, we showed that these unassisted protocols can provide significant advantage over the optimal classical strategies for the discrimination of both bosonic loss and environmental noise. In some cases, idler-free approaches can achieve performance on par with idler-assistance.

These results strongly encourage the further investigation of dynamic block protocols. Motivated by the problem setting and chosen class of probe-states, we were able to engineer high performance, quantum-enhanced unassisted protocols. However, determining the optimal unassisted protocol for specific multi-channel discrimination tasks is now an open question. It is of interest to explore more sophisticated versions of these protocols based on the optimisation of probing configurations over specific image spaces, and adaptive protocols that modify probe-domain distributions on the fly. Such studies could reveal high performance, unassisted discrimination strategies tailored to realistic applications.\par

Since this research was conducted in the CV picture based on Gaussian entangled states, this makes it particularly relevant to near term quantum sensing applications. The exploration of alternative entangled probe states is an immediate path of interest, as the employment of popular non-Gaussian entangled states could provide further enhancements to these unassisted protocols. Furthermore, the translation of this research for finite-dimensional channels is also an important topic, in which similar unassisted protocols may display strong quantum advantage.

Investigating the space of unassisted discrimination protocols is of importance for near term quantum technologies. The insights and results of this work significantly loosen the resource constraints on realisable quantum technologies that rely on pattern recognition, emphasising that idler-assistance is not always a necessity. 

\section*{Methods}

\subsection*{Quantum Barcode Decoding}

The most general pattern recognition task for binary channel patterns is quantum barcode decoding \cite{PatternRecog}. This is a multi-hypothesis discrimination task of identifying a channel pattern from the entire space of binary channel patterns. For $m$-length quantum multi-channels this is characterised by an image space ${\mc{U}_m = \{\bs{i}_1, \ldots, \bs{i}_{2^m}\}}$ which contains exactly $2^m$ possible patterns. For example, if $m=2$, the complete binary image space is explicitly ${\mc{U}_2 = \{\{B,B\}, \{B,T\}, \{T,B\}, \{T,T\} \}}$.

All other image spaces are necessarily a subset of $\mc{U}_{m}$, hence quantum barcode decoding embodies the most challenging multi-channel discrimination problem in this setting. It represents a scenario in which we have no \textit{a priori} information that can narrow the space of potential quantum channel patterns, and all $\bs{i} \in \mc{U}_m$ must be considered within the ensemble.

\subsection*{Channel Position Finding}
The task of Channel Position Finding (CPF) describes the multi-hypothesis discrimination task of locating a single target channel $\mc{E}_{T}$ hidden amongst an array of background channels $\mc{E}_B$. An $m$-channel CPF problem is associated with the image space $\mc{U}_{\text{\tiny CPF}}$ which is the set of all $m$-length multi-channels which contain exactly one target channel. \par
Let us define a function which constructs an $m$-length channel pattern with one target channel in the $x^{\text{th}}$ position of the set
\begin{equation}
e_m(x) = P_{1:x} \>\{T, \underbrace{B,\ldots, B}_{m-1 \text{ times}}\} .
\end{equation}
Here $P_{1:x}$ is a permutation operator that swaps the position of the first label $T$ with the $x^{\text{th}}$ element in the set. Then we can construct the CPF image space for $m$-modes,
\begin{equation}
\mc{U}_{\text{\tiny CPF}} = \{e_m(1), \ldots , e_m(m)\} =  \bigcup_{x=1}^m \{ e_m(x)\}.
\end{equation} 
For a $m$-channel CPF problem, $|\mc{U}_{\text{\tiny CPF}}| = m$. For example, if $m=3$, then \begin{equation}
{\mc{U}_{\text{\tiny CPF}} = \{ \{B,B,T\}, \{B,T,B\}, \{T,B,B\}\}}.
\end{equation}

More generally, we may investigate $k$-CPF, where the number of targets channel that occur within each channel pattern is precisely $k < m$, hidden amongst $m-k$ background channels. We denote this image space $\mc{U}_{\text{\tiny CPF}}^k$. Let us define a more general function which generates an $m$-length channel pattern with precisely $k$-target labels in the positions indicated by the unique indices $x_1, x_2, \ldots, x_k$,
\begin{equation}
e_m^k(x_1,\ldots,x_k) = P_{1\ldots k: x_1\ldots x_k} \{\underbrace{T,\ldots, T}_{k \text{ times}}, \underbrace{B,\ldots, B}_{m-k \text{ times}}\}.
\end{equation}
Here, each permutation operator $P_{1\ldots k:x_1\ldots x_k}$ swaps all of the target channel labels from positions $1,\ldots,k$ with the channel labels at the positions $x_1,\ldots, x_k$. Then we can construct any $k$-CPF image space by iterating over all unique permutations of the target channel labels,
\begin{equation}
\mc{U}_{\text{\tiny CPF}}^k = \bigcup_{1\leq x_1 \neq x_2 \neq \ldots \neq x_k \leq m } \{ e_m^k(x_1,\ldots,x_k)\}.
\end{equation}
For an $m$-channel k-CPF problem, $\mc{U}_{\text{\tiny CPF}}^k$ contains exact $C_m^k$ channel patterns, where $C_m^k = {m!}/({k!(m-k)!})$ is the binomial coefficient. For example, if $m=3, k=2$, then the image space is 
\begin{equation}
\mc{U}_{\text{\tiny CPF}}^2 = \{ \{T,T,B\}, \{T,B,T\}, \{B,T,T\}\}.
\end{equation}
Clearly when $k=1$ we regather the previous single CPF image space.\par

Both CPF and $k$-CPF find a number of fundamental settings within target-detection, quantum enhanced classical data-readout and environment localisation. They provide a valuable platform for studying multi-channel discrimination; if we can understand how to attain quantum enhancements in the readily analysable CPF framework, then we can learn to extract and apply these enhancements in more complex settings.

\subsection*{Bosonic Gaussian Channel Patterns}

Under the action of a single-mode GPI quantum channel, an input Gaussian state described completely via its covariance matrix (CM) $V_{\text{in}}$ with zero first moments undergoes the transformation
\begin{equation}
V_{\text{in}}~{\rightarrow}~V_{\text{out}} = (\sqrt{\tau} I) V (\sqrt{\tau} I)^T + \nu I,
\end{equation}
where $I$ is a $2\times 2$ identity matrix. The overall quantum channel can be denoted as $\mc{E}_{\tau,\nu}$ and is defined with respect to a transmissivity parameter ${0 \leq \tau \leq 1}$ describing attenuation/amplification properties and an induced noise parameter $\nu \geq 0$. \par
Binary GPI channel patterns then consist of a sequence of $m$-GPI channels with unique target/background transmissivities $\tau_B, \tau_T$ and noise properties $\nu_B, \nu_T$. Consider now a $m$-mode Gaussian state with CM $V_{\text{in}}$ and zero first moments. Let the following be a matrix function of a general variable $x$ which depends on a position $k$ in a channel pattern $\bs{i}$,
\begin{align}
I_{[x]_{\bs{i}}} &\defeq x_{i_1} I \oplus \ldots \oplus x_{i_m} I =  \bigoplus_{k=1}^m  \begin{pmatrix} x_{{i_k}} & 0 \\ 0 & x_{{i_k}} \end{pmatrix}.
\end{align}
Then a multi-mode Gaussian state which is transformed according to a GPI binary channel pattern $\rho \mapsto \mathcal{E}_{\bs{i}} (\rho)$ undergoes the following transformation on its CM in phase space
\begin{equation}
V_{\bs{i}} =( I_{[\sqrt{\tau}]_{\bs{i}}}) V_{\text{in}}  ( I_{[\sqrt{\tau}]_{\bs{i}}})^T +  I_{[\nu]_{\bs{i}}} \label{eq:CMtrans}.
\end{equation}
Therefore, it is easy to study the CMs of multi-mode Gaussian probe states interacting with GPI channel patterns.

\subsection*{Bosonic CV-GHZ States}

As discussed in Results, a CV-GHZ state can be constructed as the extension of a TMSV state to many modes. Indeed, consider a CV-GHZ state  defined over an $m$-length probe-domain, $\Phi_{\{1,\ldots,m\}}^\mu$. This $m$-mode state can be completely characterised by its CM (with zero first moments) as given in Eq.~(\ref{eq:CVGHZ}).
Here we show why maximum correlations are satisfied at $c_1 = -c_2 = c_{\text{max}}$. The symplectic spectrum of the CV-GHZ state takes the form,
\begin{align}
\nu_{-} &= \sqrt{(\mu - c_1)(\mu-c_2)},\\
\nu_{+} &= \sqrt{(\mu + (m-1)c_1)(\mu+(m-1)c_2)},
\end{align}
such that $\nu_{+}$ is $(m-1)$-degenerate. In order to capture maximal correlations (at finite squeezing), we use the bona fide condition $\nu_{\pm} \geq \frac{1}{2}$ to state that 
\begin{equation}
|c| \leq \frac{\sqrt{\mu^2 - \frac{1}{4} }}{m-1}. \label{eq:c_corr}
\end{equation}
Hence, this leads to the notion of maximal symmetric correlations when the correlation parameters satisfy $c_1 = -c_2 = c_{\text{max}} = ({m-1})^{-1}{\sqrt{\mu^2 - 1/4}}$. \par
CV-GHZ states can then be readily used to construct probe states in conjunction with a probe-domain distribution, where it can be used as a building block for more general multipartite states. This construction can be equivalently represented in phase space, where tensor products over sub-states become direct sums over sub-CMs. More precisely, given an $N$-partite probe-domain distribution $\mc{S}$ (disjoint or non-disjoint) we can equivalently express the global CV-GHZ input state $\Phi_{\mc{S}}^{\mu}$ via its CM,
\begin{equation}
\Phi_{\mc{S}}^{\mu} \rightarrow V_{\mc{S}}^{\mu} = \bigoplus_{j=1}^{N} V_{\bs{s}_j}^{\mu},
\end{equation}
where $V_{\bs{s}_j}^{\mu}$ is the CM of a $|\bs{s}_j|$-mode CV-GHZ state irradiated over the modes contained in the probe-domain $\bs{s}_j$.

\subsection*{Numerical Computation of Error Bounds}

Consider two unique, $m$-length Gaussian channel patterns $\mc{E}_{\bs{i}}$ and $\mc{E}_{\bs{i}^{\prime}}$ which are probed by two identical, $m$-mode CV-GHZ states $\Phi_{\{1,\ldots,m\}}^{\mu}$. We can conveniently write the output states from these interactions,
\begin{align}
\Phi_{\bs{i}}^{\mu} &= \mc{E}_{\bs{i}} (\Phi_{\{1,\ldots,m\}}^{\mu} ) \rightarrow V_{\bs{i}}^{\mu},\\
\Phi_{\bs{i}^\prime}^{\mu} &= \mc{E}_{\bs{i}^\prime} (\Phi_{\{1,\ldots,m\}}^{\mu} )\rightarrow V_{\bs{i}^\prime}^{\mu}.
\end{align}

Now consider the fidelity between these two output states $F(\Phi_{\bs{i}}^{\mu}, \Phi_{\bs{i}^{\prime}}^{\mu})$. Thanks to the Gaussianity of CV-GHZ states and GPI multi-channels, the fidelity between these states can be computed exactly using only their phase space representations using the formulae from \cite{GFid},
\begin{equation}
F(\Phi_{\bs{i}}^{\mu}, \Phi_{\bs{i}^{\prime}}^{\mu}) = F_G(V_{\bs{i}}^{\mu}, V_{\bs{i}^{\prime}}^{\mu}),
\end{equation}
where we denote $F_G$ as the Gaussian fidelity function. \par

In summary, we have a way to represent the input probe states of unassisted block protocols, through $V_{\mc{S}}^{\mu}$; the ability to describe output states by transforming input states according to GPI multi-channels $\mc{E}_{\bs{i}}$ as in Eq.~(\ref{eq:CMtrans}), 
\begin{equation}
\Phi_{\mc{S},\bs{i}}^{\mu} = \mc{E}_{\bs{i}}(\Phi_{\mc{S}}^{\mu} ) \rightarrow V_{\mc{S},\bs{i}}^{\mu},
\end{equation}
and the means to compute the fidelity between any two output states. Given these techniques, and a multi-channel discrimination problem $\{\pi_{\bs{i}} ; \mc{E}_{\bs{i}} \}_{\bs{i}\in\mc{U}}$, we can readily compute the fidelity-based error probability lower and upper bounds,
\begin{align}
p_{\text{err}} &\geq \frac{1}{2} \sum_{\bs{i}\neq\bs{i}^{\prime} \in \mc{U}} \pi_{\bs{i}} \pi_{\bs{i}^{\prime}} F_G^{2M}\big[ V_{\mc{S},\bs{i}}^{\mu}, V_{\mc{S},\bs{i}^{\prime}}^{\mu},\big] ,\label{eq:GLB}\\
p_{\text{err}} &\leq \sum_{\bs{i}\neq\bs{i}^{\prime} \in \mc{U}} \sqrt{\pi_{\bs{i}} \pi_{\bs{i}^{\prime}}} F_G^{M}\big[ V_{\mc{S},\bs{i}}^{\mu}, V_{\mc{S},\bs{i}^{\prime}}^{\mu}, \big]. \label{eq:GUB}
\end{align}\par
To study the error bounds of dynamic block protocols, we need only invoke the dynamic/fixed protocol transformation discussed in the Results section. In this way, we modify the channel patterns according to the probe-domain distribution $\bs{i}\rightarrow \bs{\nu_i}$. By computing the fidelity between the outputs of CV-GHZ states irradiated over the modified patterns 
\begin{equation}
F_G \big[ V_{\mc{S},{\bs{\nu_i}}}^{\mu}, V_{\mc{S}, {\bs{\nu}_{\bs{i}^{\prime}}}}^{\mu} \big],
\end{equation}
error bounds can be readily computed for dynamic block protocols. The numerical methods presented here can always be used for fixed or dynamic block protocols, and more generally using \textit{any} Gaussian input states.

\subsection*{Classical Performance}
Here we collect expressions for the classical fidelities using optimal coherent states for the multi-channel discrimination settings explicitly studied in this work. These can then be used to derive exact error-bounds, and benchmark quantum advantage. \par
The best classical protocol for discriminating a single pure-loss channel is achieved by a block protocol using coherent states. Indeed, the optimal (energy constrained) $M$-copy, single-mode coherent state given by \cite{PatternRecog},
\begin{equation}
\alpha_{\text{coh}}^{\otimes M} =  \ket{{M N_S}}\!\bra{M N_S},
\end{equation}
where $N_S$ is the mean photon number of the signal state. If a single pure-loss channel possesses transmissivity $\eta_B$ or $\eta_T$, the fidelity between the two possible single-copy output states is given by
\begin{equation}
F_{\text{cl}}^{\text{loss}}= \exp\left[-\frac{N_S}{2}(\eta_B - \eta_T)^2\right].
\end{equation}
For $m$-length multi-channels, we simply use $m$ single mode coherent states to discriminate each channel independently.\par
For additive noise channels, $\tau=1$ and the task is to discriminate between background/target noise $\nu_{B},\nu_T > 0$. The optimal classical input state is just the $m$-copy vacuum state $\ket{0}^{\otimes m}$, since displacements or phase shifts have no impact on the output states from the channel. The single probe copy fidelity between the potential single-mode output states can be written as \cite{OptEnvLoc},
\begin{equation}
F_{\text{cl}}^{\text{add}} = \frac{1}{\sqrt{(\nu_T +1)(\nu +1)} - \sqrt{\nu_T \nu_B}}.
\end{equation}
Using these fidelity expressions within the error probability bounds from Eqs.~(\ref{eq:LB}) and (\ref{eq:UB}), we can provide effective classical benchmarks for both multi-channel discrimination settings. \\

\subsection*{Idler-Assisted Performance}
We can similarly collect expressions for idler-assisted block protocols in the context of bosonic pure-loss channel patterns and environment localisation.\par
For the discrimination of bosonic loss, one can employ an idler-assisted protocol in which each channel is probed with one mode from an $M$-copy TMSV state, $\Phi^{\mu}$, where $\mu = N_S + 1/2$ is the level squeezing. Consider a single pure-loss channel $\mc{E}_{j}$ which may have transmissivity $\eta_j \in \{\eta_B,\eta_T\}$. While one mode interacts with the channel, the other mode is perfectly protected (in a quantum memory for instance) and thus undergoes the action of an identity channel. The output state is then simply the finite-energy Choi state ${\Phi_{\mc{E}_j}^{\mu} = \mc{E}_j\otimes\mc{I}(\Phi^{\mu})}$. The fidelity between the output states $\Phi_{\mc{E}_B}^{\mu}$ and $\Phi_{\mc{E}_T}^{\mu}$ is \cite{PatternRecog}
\begin{equation}
F_{\text{idler}}^{\text{loss}} = \frac{1}{1 + N_S \Delta },
\end{equation}
where ${\Delta = 1-\sqrt{(1-\eta_B)(1-\eta_T)} - \sqrt{\eta_B\eta_T}}$. By extending this to $m$-channels using $M$-copy probes, we can easily bound performance of the idler-assisted block protocol.\par

We can perform an identical analysis for environment localisation by computing the fidelity between possible output states of a pair of additive-noise channels with target noise or background noise $\nu_B$/$\nu_T$. In this case, it is convenient to utilise the parameter $\mu = N_S+{1}/{2}$, where the output fidelity reads \cite{ThermalPatt}
\begin{equation}
F_{\text{idler}}^{\text{add}} = \frac{2\mu \sqrt{\nu_T\nu_B} + \sqrt{(2\mu \nu_T+1)(2\mu \nu_B+1)}}{2\mu (\nu_T+\nu_B)+1}.
\end{equation}
Once again extending these fidelities to consider $m$-channels and using $M$-copy probes states, we can easily bound the performance of idler-assisted block protocols.\\

\subsection*{Fidelity  Degeneracies of CV-GHZ States}

Consider an arbitrary, $N$-element image space $\mc{U} = \{ \bs{i}_1, \bs{i}_2,\ldots,\bs{i}_N\}$  which generate $m$-length quantum channel patterns, and the associated multi-channel ensemble $\{ \pi_{\bs{i}} ; \mc{E}_{\bs{i}}\}_{\bs{i}\in\mc{U}}$. Now, consider the use of an unassisted block protocol using $m$-mode CV-GHZ states. As we know from Results, we can benchmark the performance of this protocol via the fidelity-based error probability bounds in Eqs.~(\ref{eq:LB}) and (\ref{eq:UB}).

While this can be achieved numerically, it can become inefficient. The total number of ways that we can choose unequal pairs of channel patterns is $N(N-1)$. This means that in general, there exist $N(N-1)$ potentially unique, non-unit fidelities that one needs to compute in order to calculate the error bounds. For large pattern spaces this can be enormous, making it difficult to analytically keep track of all possible output fidelities, or numerically perform these sums. 

However, thanks to their symmetry, when using CV-GHZ states as quantum probes the number of unique fidelities that may occur is dramatically reduced. The CV-GHZ state symmetry causes many of the unique output fidelities within Eqs.~(\ref{eq:LB}) and (\ref{eq:UB}) to be highly degenerate. Indeed, fidelity degeneracy tells us that if there are exactly $g_{\text{fid}}$ unique output fidelities, typically $g_{\text{fid}} \ll N(N-1)$.

Let us be more precise: Consider a pair of image spaces of $m$-length, binary channel patterns: one is the $k$-CPF image space $\mc{U}_{\text{CPF}}^k$, such that each pattern contains precisely $k$-target channels, and the other is a $l$-CPF image space $\mc{U}_{\text{CPF}}^l$ such that each pattern contains precisely $(l\neq k)$-target channels. Take two identical $m$-mode CV-GHZ states $\Phi_{\{1,\ldots,m\}}^{\mu}$ which interact with the multi-channels $\mc{E}_{\bs{i}}$ and $\mc{E}_{\bs{i}^{\prime}}$, resulting in two unique output states $\Phi_{\bs{i}}^{\mu}$ and $\Phi_{\bs{i}^{\prime}}^{\mu}$. Now consider the fidelity,
\begin{equation}
F(\Phi_{\bs{i}}^{\mu}, \Phi_{\bs{i}^{\prime}}^{\mu}), \text{ for } \bs{i} \in \mc{U}_{\text{CPF}}^k,\>\bs{i}^{\prime} \in \mc{U}_{\text{CPF}}^l.
\end{equation}
We find that this fidelity is equivalent for all pairs of channel patterns $\bs{i}, \bs{i}^{\prime}$ which have the same Hamming distance. That is, for all
\begin{equation}
\bs{i} \in \mc{U}_{\text{CPF}}^k,\>\bs{i}^{\prime} \in \mc{U}_{\text{CPF}}^l, \text{ s.t } \text{Hamming}(\bs{i},\bs{i}^{\prime}) = d > 0,
\end{equation}
the fidelity $F(\Phi_{\bs{i}}^{\mu}, \Phi_{\bs{i}^{\prime}}^{\mu})$ is completely degenerate. For a rigorous proof of this, please see \cite{ABDMCD}. Fidelity degeneracies are extremely useful, and can help to not only improve numerical efficiency, but reveal analytical insights.

\begin{figure}[t!]
\centering
\hspace{-0.5cm} \includegraphics[width=0.87\linewidth]{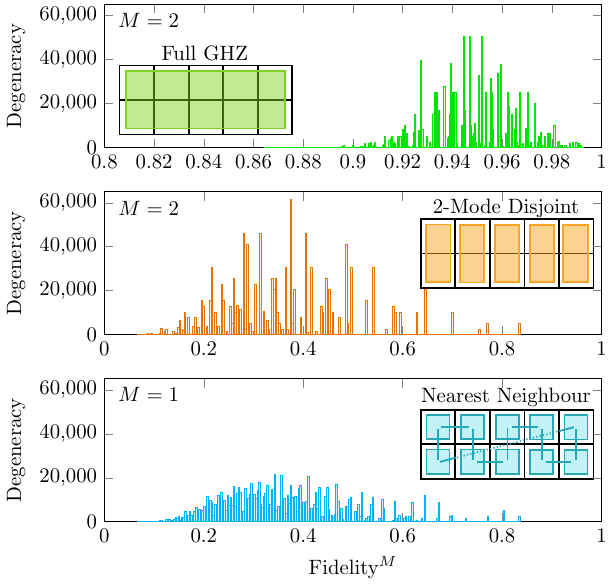}\\
\caption{\textbf{CV-GHZ Fidelity Degeneracies}: Histograms of fidelity degeneracies for unequal $m=10$ pure-loss channel patterns over the complete image space of binary channel patterns $\>\mathcal{U}_{10}$, for a selection of probe-domain distributions: $m$-mode  CV-GHZ state (green), disjointly distributed TMSV states (orange) and nearest neighbour distributed TMSV states (blue). The fixed block protocols are using $M=2$ probe-copies, while the dynamic nearest neighbour protocol is using $M=1$ copies. In this way, the average channel use of all the protocols is the same, $\bar{M}=2$.}  
\label{fig:Degen}
\end{figure}

In Fig.~\ref{fig:Degen} we have numerically investigated the fidelity degeneracy properties of a number of different unassisted dynamic/disjoint protocols for the discrimination of pure-loss channel patterns. Here we observe two clear points; CV-GHZ states lose distinguishability when we widen their domain size as expected, due to weakening quantum correlations (discussed in the Results). This can be seen by comparing the output fidelity spectrum of the $m=10$ mode CV-GHZ probe state (green) compared to the other probe-domain distributions.

Furthermore, non-disjoint probe domain distributions are able to ``spread out" the degeneracies involved with disjoint probing protocols. In Fig.~\ref{fig:Degen}, we compare a fixed block protocol using a disjoint, exclusively two-mode distribution of probe domains (orange). We then use a dynamic nearest neighbour protocol with the exact same number of probe modes (blue). While the output fidelity distributions possess a similar spread in values, the variation in probe-domains raises many of the degenerate fidelities. In doing so, it flattens the overall distribution, and gives rise to more distinguishable output fidelities.

\subsection*{Fidelity Degeneracies of TMSV states}

We will now focus on the case of using $m=2$ CV-GHZ states, i.e.~TMSV states. As discussed in Results, these states maximise their entanglement content with respect to input energy, since quantum correlations do not need to be spread across many modes. Furthermore, they offer the simplest test case for analytically investigating fidelity degeneracies. This will help to unveil concrete reasons for the discrepancy between fixed and dynamic protocols.

We wish to identify all the possible, unique output fidelities associated with TMSV states irradiated over $m=2$ length binary channel patterns. We can summarise this image space easily as it is very small,
\begin{equation}
\mc{U}_{2} = \{\{B,B\},\{B,T\}, \{T,B\}, \{T,T\}\}.
\end{equation}
Thanks to the fidelity degeneracy properties discussed in the previous section it turns out that there are only four unique \textit{sub}-fidelities that can occur when one irradiates two-mode binary channel patterns with unassisted TMSV states. Here, we define a sub-fidelity as a single output fidelity that occurs between specific pairs of channel patterns. These sub-fidelities are completely determined by the number of the target channels, $k$ and $l$, contained within the considered channel pair, $\bs{i}$ and $\bs{i}^{\prime}$ respectively. Hence we will denote each sub-fidelity in the form $F_{k:l}$ where $k$ ($l$) indicates the number of target channels in the channel pattern $\bs{i}$ ($\bs{i}^{\prime}$). Doing so, we can write all the unique, two-mode sub-fidelities
\begin{align}
\begin{aligned}
&F_{0:1}(\mu) ,  \text{ when } \bs{i} = \{ B,B\}, \bs{i}^{\prime} \in \{ \{T,B\},\{B,T\}\},  \\
&F_{0:2}(\mu) , \text{ when } \bs{i} = \{ B,B\}, \bs{i}^{\prime} = \{T,T\},  \\
&F_{1:1}(\mu) , \text{ when } \bs{i} = \{ B,T\}, \bs{i}^{\prime} = \{T,B\},  \\
&F_{1:2}(\mu) , \text{ when } \bs{i} = \{ T,T\}, \bs{i}^{\prime} \in \{ \{T,B\},\{B,T\}\}.  
\end{aligned}
\label{eq:TMSubs}
\end{align}
The Bures fidelity is a symmetric function, therefore the order of $\bs{i}$ and $\bs{i}^{\prime}$ is irrelevant. \par
These are the only fidelities that can occur when using TMSV states over pairs of $m=2$ length channel patterns.
Furthermore, the fidelity is multiplicative, meaning that 
\begin{equation}
F(\rho\otimes\rho^{\prime},\sigma\otimes\sigma^{\prime}) =  F(\rho,\sigma) \cdot F(\rho^{\prime},\sigma^{\prime}).
\end{equation}
Hence, when using collections of exclusively two-mode states following some probe-domain distribution (such as in the nearest neighbour protocol), then all of their unique output fidelities will always be a specific product of these sub-fidelities in Eq.~(\ref{eq:TMSubs}). Hence, these sub-fidelities can be used to completely characterise any unassisted discrimination protocol using exclusively TMSV states (see Ref.~\cite{ABDMCD} for more details).

While this may seem like an unnecessary level of detail, the investigation of these sub-fidelities helps to reveal critical features of dynamic block protocols. Each of the sub-fidelities in Eq.~(\ref{eq:TMSubs}) is a unique function that can be analytically characterised via the Gaussian fidelity formulae from \cite{GFid}. They each possess a unique behaviour dependent upon the multi-channel discrimination setting that we are considering. If one of the sub-fidelities is typically very large, this means that the specific pair of channel patterns that is refers to are very difficult to discriminate. If a sub-fidelity is very small, then the pair of channels it refers to are very easy to discriminate.
For example, if $F_{0:2} \gg F_{1:1}$ in a particular problem setting, then it is much easier to discriminate the patterns $\{ B,T\}$ from $\{T,B\}$, rather than $\{ B,B\}$ from $\{T,T\}$.

Most importantly, when utilising unassisted discrimination protocols, there is \textit{an inconsistency of distinguishability} between different collections of quantum channels. This inconsistency leads to the corrective behaviour that dynamic protocols can provide. We will convey this inconsistency by considering the settings studied in the Results section.

\begin{figure}[t!]
\includegraphics[width=\linewidth]{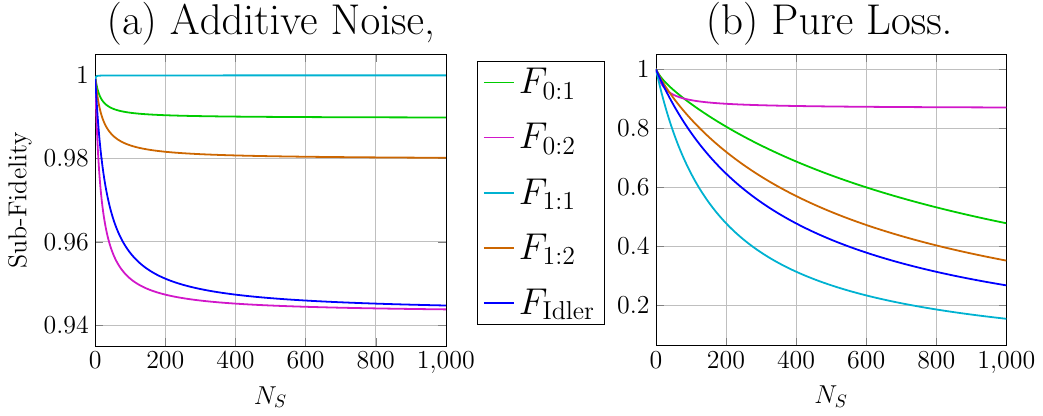}
\caption{\textbf{TMSV Sub-fidelities}: Two mode sub-fidelity behaviour with respect to increasing signal energy $N_S$ for unassisted, single-copy TMSV states interacting with $m=2$ length channel patterns for  (a) additive noise channels $\nu_T,\nu_B = 0.01,0.02$ and (b) pure-loss channels $\eta_T,\eta_B = 0.97,0.99$. }
\label{fig:SubFids}
\end{figure}

\subsection*{Analytical Insight for Dynamic Protocols}

Let us take the example of environment localisation for $m=2$ length binary channel patterns of additive-noise channels with target noise $\nu_T$ and background noise $\nu_B$.  When we look closely at the sub-fidelities concerned with this setting, we notice a glaring inconsistency. 
Three of these sub-fidelities ($F_{0:1}$, $F_{0:2}$ and $F_{1:2}$) all assume their minimum in the limit of infinite squeezing, 
\begin{equation}
\text{min}~F_{k:l}(\mu) = \lim_{\mu \rightarrow \infty} F_{k:l}({\mu}).
\end{equation}
That is, by increasing the energy of our input states, we can expect to improve the discrimination of the appropriate pairs of channel patterns. This sub-fidelity behaviour is displayed in Fig.~\ref{fig:SubFids}(a).

However, this is not the case for the sub-fidelity $F_{1:1}({\mu})$, which is concerned with the discrimination of the pattern $\{B,T\}$ from $\{T,B\}$ (and vice versa). This sub-fidelity explicitly takes the form
\begin{align}
&F_{1:1}(\mu) = \frac{1}{\theta - \sqrt{\xi_- \xi_{+}}},
\end{align}
where we define the quantities
\begin{align}
&\theta = 2 \nu_B \nu_T+1 + 2 \mu (\nu_B+\nu_T),\\
&\xi_{\pm} = \theta - 1 \pm (\nu_B - \nu_T).
\end{align}
In the limit of infinite squeezing, we find that 
\begin{equation}
\lim_{\mu \rightarrow \infty} F_{1:1}({\mu}) = 1,
\end{equation}
meaning that in the limit of infinite probe state energy, the pair of channel patterns $\{B,T\}$ and $\{T,B\}$ become completely indistinguishable. Clearly, this will have a hugely detrimental effect on discrimination performance which cannot be remedied by increasing the input probe energy.

A similar effect can be observed within pure-loss channel patterns. Let us consider the case of probing $m=2$ length binary channel patterns such that each channel is a pure-loss channel with either a target transmissivity $\eta_T$ or background transmissivity $\eta_B$. Now, we find that the  two mode sub-fidelities $F_{0:1}$, $F_{1:1}$, and $F_{1:2}$ tend to zero in the limit of infinite squeezing,
\begin{equation}
\text{min}~F_{k:l}(\mu)  = \lim_{\mu \rightarrow \infty}  F_{k:l}({\mu}) = 0.
\end{equation}
This means that in the limit of infinite energy and maximum entanglement, the channel pairs that characterise each of these sub-fidelities become perfectly distinguishable.

However, we may focus on the quantity $F_{0:2}({\mu})$ which defines the distinguishability of the patterns $\{B,B\}$, $\{T,T\}$. This sub-fidelity takes a relatively compact form given by,
\begin{align}
&F_{0:2}(N_S)  = \frac{ 2N_S\sqrt{\kappa_1}  + \sqrt{\kappa_2}}{{1 - N_S(\eta_B+\eta_T-2)(\eta_T+\eta_B)}}.
\end{align}
where we use $\mu= N_S + {1}/{2}$ as before and we define the quantities
\begin{align}
\kappa_1 &= \eta_B\eta_T(\eta_T-1)(\eta_B-1), \\
\kappa_2 &= 1 - N_S(\eta_B + \eta_T - 2)(\eta_B +\eta_T) - 4N_S^2\kappa_1.
\end{align}
In the limit of infinite squeezing, this sub-fidelity $F_{0:2}$ instead finds the finite quantity
\begin{equation}
\lim_{\mu \rightarrow \infty} F_{0:2} = 4\sqrt{\frac{ \eta_B \eta_T (\eta_B-1) (\eta_T-1)}{(\eta_B+\eta_T-2)^2 (\eta_B+\eta_T)^2}}, 
\end{equation}
which is non-zero when either $\eta_{j} \neq 1$. Therefore, even when using infinitely squeezed input states, the patterns $\bs{i} = \{B,B\}$ and $\bs{i}^{\prime} = \{T,T\}$ are not perfectly distinguishable via unassisted input states. The behaviour of these sub-fidelities are displayed in Fig.~\ref{fig:SubFids}(b).

It is now clear why dynamic protocols are so effective at redeeming the performance of these multi-channel discrimination tasks. When using unassisted, multipartite entangled probe states, the distinguishability of the output states can vary considerably, dependent upon the collection of quantum channels that they interact with. By overlapping probe-domains, channels can be probed in conjunction with different collections of channels in the pattern. In doing so, we are increasing the likelihood of probing a more a distinguishable collection. These more distinguishable channel regions are then able to \textit{correct} the errors invoked by probes interacting with poorer regions.

As an example, let us take the discrimination of Gaussian additive-noise channels with TMSV states according to some probe-domain distribution. Consider an $m=4$ length channel pattern, 
\begin{equation}
\bs{i} = \{i_1,i_2,i_3,i_4\} = \{ B, T, T, B\}, 
\end{equation}
that we wish to discriminate. Here, we first consider a fixed block protocol that follows the probe-domain distribution $\mc{S}_{\text{d}} = \{ \{1,2\},\{3,4\} \}$. As a result of this distribution, we will possess the following TMSV states which irradiate specific channel sub-patterns,
\begin{gather}
\begin{gathered}
\Phi_{\{1,2\}}^{\mu} \text{ irradiates } \{i_1,i_2\} = \{B,T\},\\
\Phi_{\{3,4\}}^{\mu} \text{ irradiates } \{i_3,i_4\} = \{T,B\}.
\end{gathered}
\end{gather} 

As discussed earlier in this section, the sub-fidelity $F_{1:1}$ is very poor, meaning that the sub-patterns $\{ B,T\}$ and $\{T,B\}$ are very difficult to distinguish from one another. Therefore, if we irradiate a sub-pattern $\{ B,T\}$ or $\{T,B\}$ directly with a TMSV state, our overall discrimination ability will be very ineffective, as we will struggle to determine which is the true pattern. It is highly desirable to avoid instances of this kind of pattern interaction, but obviously we cannot know prior to interaction where these pairs of channel patterns arise (this would defeat the purpose of discrimination). The embodies a critical weakness of fixed block protocols. If we unwittingly choose a probe-domain distribution which irradiates input states over poorly distinguishable collections of patterns, the ability to discriminate the overall channel pattern will be compromised. 

We now see why varying the probe-domains is so effective. Consider a non-disjoint probe-domain distribution, $\mc{S}_{\text{nd}} = \{ \{1,2\},\{2,3\},\{3,4\},\{1,4\} \}$ (this is in fact the nearest neighbour protocols discussed in Results). Consider the same  $m=4$ length channel pattern, $\bs{i} = \{ B, T, T, B\}$. As a result of this distribution, we will possess the following TMSV states which irradiate the specific channel sub-patterns,
\begin{gather}
\begin{gathered}
\Phi_{\{1,2\}}^{\mu} \text{ irradiates } \{i_1,i_2\} = \{B,T\},\\
\Phi_{\{2,3\}}^{\mu} \text{ irradiates } \{i_2,i_3\} = \{T,T\},\\
\Phi_{\{3,4\}}^{\mu} \text{ irradiates } \{i_3,i_4\} = \{T,B\},\\
\Phi_{\{1,4\}}^{\mu} \text{ irradiates } \{i_1,i_4\} = \{B,B\}.
\end{gathered}
\end{gather} 

While we are still unfortunately irradiating the poorly distinguishable collections of channels $\{i_1,i_2\} = \{ B,T\}$ and $\{i_3,i_4\} = \{T,B\}$ with two of our input probes, we now also apply probe states to the sub-patterns
$\{i_2,i_3\} = \{T,T\}$ and $\{i_1,i_4\} = \{B,B\}$. These collections of channels are much more distinguishable, and invoke the stronger sub-fidelities $F_{0:1}$, $F_{0:2}$, and $F_{1:2}$. Hence, by varying our probe domains we are able to gather different, more distinguishable ``opinions" of regions of the channel pattern. The stronger distinguishability of the regions $\{i_2,i_3\}$ and $\{i_1,i_4\}$ can help to correct the contribution of the poorly distinguishable channel collections.

It is important to note that this improved performance is not connected to an increased number of probe modes. Recall that we can fairly compare dynamic/fixed block protocols with the same average channel use $\bar{M}$. With equivalent resources, the dynamic protocol will outperform the fixed version thanks to its variable probe-domains.

\begin{acknowledgments}
C.H acknowledges funding from the EPSRC via a Doctoral Training Partnership (EP/R513386/1). S.P acknowledges funding from the European Union’s Horizon 2020 Research and Innovation Action under grant agreement No.~862644 (Quantum readout techniques and technologies, QUARTET). 
\end{acknowledgments}

%\bibliography{IF_Bib}


\end{document}